\pdfoutput=1
\documentclass[prl,twocolumn,superscriptaddress,showpacs,floatfix]{revtex4-1}
\usepackage{epsfig}
\usepackage{color}
\usepackage{graphics}
\usepackage{amsmath}
\usepackage{amssymb}
\usepackage{braket}

\newcommand{\Aotz} {z^{x_1}_{1}z^{x_2}_2}

\newcommand{\Atoz} {S_{21}z^{x_2}_{1}z^{x_1}_2}

\newcommand{\CR}{{\bf C}}
\newcommand{\CT}{{\bf T}}
\newcommand{\CS}{{\bf S}}
\newcommand{\CSt}{{\bf S^+}}
\newcommand{\CJ}{{\bf J}}
\newcommand{\Ld}{{\bf L_{1d}}}
\newcommand{\q}{{\it Q1D}}
\newcommand{\Qa}{{\it Q_{{\bf CT}}}}
\newcommand{\Qb}{{\it Q_{{\bf TS}}}}
\newcommand{\Qc}{{\it Q_{{\bf SJ}}}}

\begin{document}

\title{A Kinetically Constrained Model with a Thermodynamic Flavor}
\author{S. S. Ashwin}
\email{ss.ashwin@gmail.com}
\affiliation{Department of Applied Physics, Nagoya University,
Nagoya, Japan}


\date{\today}

\begin{abstract}
Kinetically constrained models (KCM) generically have trivial thermodynamics and yet manifest rich glassy dynamics.  In order to resolve the thermodynamics-dynamics disconnect in KCMs, we derive a KCM by coarse-graining a non-trivial thermodynamic model with a solvable partition function. Blending of thermodynamics to the KCM makes the role of landscape properties on the relaxation times, fragility and its crossover analytically transparent. Using Bethe ansatz, we calculate the time-dependent $N$-point probability function in the distinct dynamical regimes of the landscapes; the contribution of the glassy term is identified. 
\end{abstract}

\maketitle
When an equilibrium fluid avoids crystallization, it enters a metastable phase. Within a small range of volume-fraction, the relaxation times in the metastable phase increases over decades in magnitude. Eventually, when the relaxation times of the fluid tends to become comparable to experimental timescales, the fluid is defined to be glass.
The interplay of thermodynamics and the increasingly important role of core interactions, leading to kinetic constraints, makes the mechanism of glass formation complex. 
Understanding the precise relationship between dynamics and thermodynamics in the metastable phase of fluid has thus been a long-standing challenge to statistical physics. 
Thermodynamics-based theories such as the Adam-Gibbs model ~\cite{AdamGibbs1965} and RFOT ~\cite{Lubchenko2007,Bouchaud2004,KirkpatrickRMP} do provide a relationship between relaxation time and thermodynamics in glass-formers. 
Unfortunately, such approaches subsume all the kinetic information into the thermodynamics, making thermodynamics-dynamics relationship obscure.

Alternately, kinetically constrained models (KCM) ~\cite{Ritort2003,Garrahan2010} describe slow dynamics incorporating restrictions arising from either core interactions or some coarse-graining of these interactions in glassformers. KCM kinetics are typically represented on a lattice using simple rules emerging from local constraints. KCMs generically have trivial thermodynamics, yet give rise to rich glassy dynamics independent of thermodynamic features. KCMs preclude thermodynamics, interpreting glassiness as a purely dynamical phenomenon.
 ~KCMs would have been excellent candidates to elucidate the relationship between dynamics and thermodynamics provided they had nontrivial thermodynamics governing them.  A paradigm shift in the study of KCMs occurred with the introduction of s-ensemble approach \cite{Garrahan2007,Garrahan2009} based on Ruelle's thermodynamic formalism~\cite{Ruelle1978, Ruelle1985}. Here, the conventional partition function is replaced by the dynamical partition function, and the glass transition is viewed as a dynamical transition involving a trajectory phase transition in space-time-\cite{Garrahan2007,Garrahan2009,Hedges2009}. Though this approach has provided new insights into the glass transition problem from a non-equilibrium viewpoint, the original problem of the connection between KCMs and thermodynamics is circumvented. In this letter, we take a step back and connect KCM kinetics to equilibrium thermodynamics.


The energy landscape/density landscape approach ~\cite{SastryNature1998, PabloNature2001} is an intuitively powerful way to understand glassy dynamics from a thermodynamic point of view. The geometry of the landscape is related to the thermodynamics of a glass-former. For example, the number of the explored basin minima on the landscape is related to the configurational entropy. Fragility~\cite{Angell1995} characterizes how rapidly a glass former approaches glass transition on cooling/compression, {\it strong} glass formers do so slowly and {\it fragile} are fast. Arrhenius dynamics characterize the strong, and non-Arrhenius, the fragile glass-former.
The landscape approach has especially provided deep insights into fragility \cite{SastryNature2001} and the fragile to strong crossover \cite{AngelaniPRL2000, BroderixPRL2000, CoslovichArxiv2018} generically seen in glass formers~\cite{CoslovichArxiv2018}. 
 Since the landscape geometry has strong links with dynamics as well \cite{PabloNature2001}, relating a KCM to thermodynamics via the landscape is a viable approach to elucidate the dynamics-thermodynamics relationship.
 
 We start with a model of a confined fluid whose partition function is calculable \cite{Barker} and exhibits several generic glassy features \cite{Bowles2000,BowlesPRE2006,MoorePRE2014} in the dense regime such as rapid increase in the relaxation time, stretched exponential relaxations~\cite{BowlesPRE2006}, fragile to strong cross-over~\cite{MahdiPRL2012,MahdiPRE2015} and dynamical heterogeneity~\cite{MoorePRE2016}. The simplicity of the confined fluid model ($\q$) clarifies the role of thermodynamics on the local coarse-grained dynamics. We identify the so-called glassy {\it onset}\cite{SastryNature1998}, its underlying geometry, and thus construct a KCM where the interplay of thermodynamics and kinetic constraints are transparent. This connection brings us to the main results of this letter: (i) We analytically calculate the relaxation time ($\tau$) of the fluid as a function of the fraction unstable saddle point modes and the inherent structure volume fraction. We also obtain the fragility index across the fragile to strong crossover. (ii) The time-dependent N-point probability function (TNPF) completely defines the physics of a liquid. Our model is amenable to a Bethe ansatz (BA) treatment, enabling us to calculate the TNPF in the distinct glassy dynamical regimes ~\cite{PabloNature2001} and we isolate the glassy contribution to the TNPF.
\begin{figure}[t]
\includegraphics[width=3.2in]{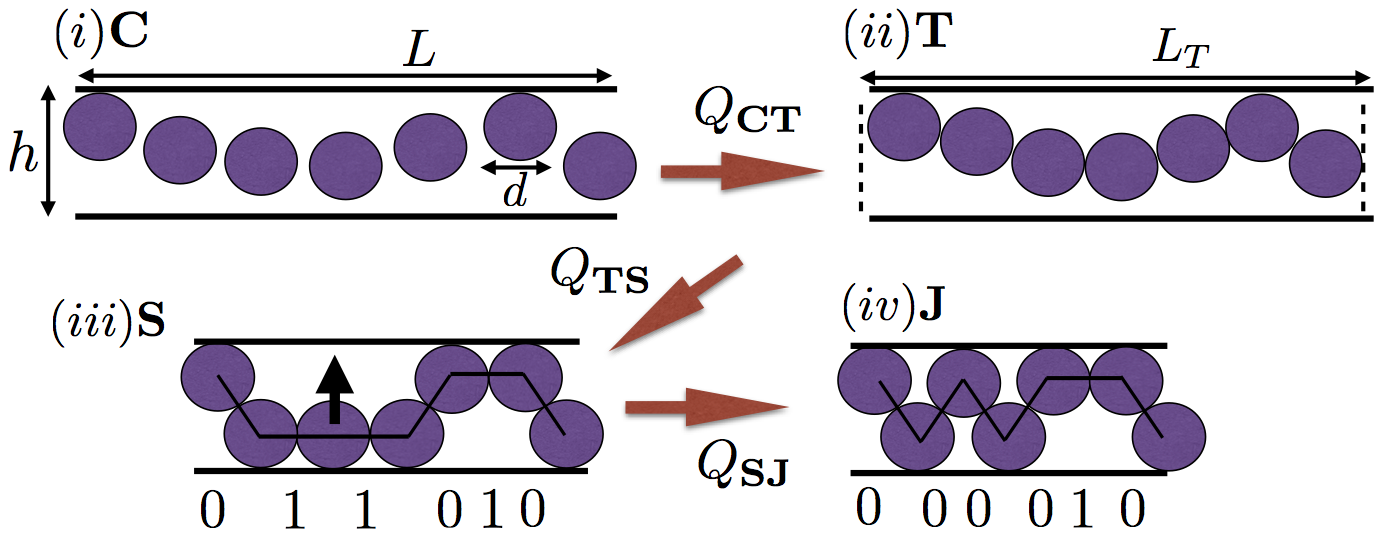}
\caption{ (i) An equilibrium configuration in $\CR$. (ii) A tangent disc configuration in $\CT$ on application of $Q_{CT}$. (iii)A saddle configuration in $\CS$ on application of $Q_{TS}$. Bonds 1,0 are indicated. Note the vertically unstable disc in the 1-1 bond, (iv) A configuration in $\CJ$, application of $Q_{SJ}$ all maps 1-1$\rightarrow$ 0-0.}  
\label{fig1}
\end{figure}

 $\q$: Consider $N$ hard disks of diameter $d=1$ confined between two hard walls of length $L$ separated by a distance $h$, where
  \begin{equation}\label{eq:1}
  h=d(1+\surd{3}/4)-\delta 
  \end{equation}
  Also, $\delta$ is such that $h=1.866d$, this geometric constraint ensures no next nearest neighbor contacts, see Fig. 1(i). The configurational space of $N$-discs is denoted by $\CR$. We use a transfer matrix approach \cite{Kofke} as it lends itself to connect the $\q$ to a KCM.  The geometric constraint in $\q$, allows one to simplify the configurational volume integral by representing it as an integral over configurational volume explored by tangent (contacting) discs. The quench protocol $\Qa$ for creating tangent discs from a liquid configuration involves translating disks in the direction parallel to the wall so that all the disks touch their nearest neighbors (See Fig. 1(ii)).  Let the total length of the tangent disc chain along the x-axis be $L_T$. The configurational space of $N$-tangent discs is $\CT$.   The region devoid of discs has a configurational volume $(L-L_T)^N$. Taking a Legendre transform results in a partition function in the $NPT$ ensemble of the form:
\begin{equation}\label{Zeqn}
Z = \frac{1}{\Lambda^{2N}(\beta P_L)^{N+1}}\int_{d/2}^{h-d/2} dy K_N(y,y)
\end{equation}
The kernel  $K_N(y',y'')=\int_{d/2}^{h-d/2} dy K(y',y)K_{N-1}(y,y'')$, where $K(y_1,y_2) =\exp(-\beta P_L \Delta_x(y_1,y_2))$, $y_1,y_2$ are the y-coordinates nearest neighbor discs in $\CT$, $\Delta_x(y_1,y_2)=\sqrt{d^2-(y_1-y_2)^2}$ and $\Lambda$ is the thermal wavelength. Here, $P_L$ is the pressure conjugate to the volume $L-L_T$. Note the configurational information in $\CT$ is encoded in $K_N$. The introduction of $\CT$ is crucial as it is the average geometry of the discs in $\CT$, which relates the partition function to the KCM dynamics.

 On squeezing the tangent disc chain along the horizontal (we call the protocol: $\Qb$) resulting in a configuration wherein all disks get jammed along the horizontal (Fig. 1(iii)), the configuration space of these quenched configurations is $\CS$. The geometric constraint~(\ref{eq:1}) ensures that configurations in  $\CS$ have precisely two kinds of local bonds (i) bond 1: disks are horizontally adjacent and (ii) bond 0: disks are diagonally adjacent~\cite{Bowles2000, BowlesPRE2006, Ashwin2009}. 
An adjacent pair of bonds involves three disks. The central disk participating the bond pairs 1-0, 0-1 and 0-0  is jammed. In contrast, the central disk in the 1-1 bond pair (shown in Fig 1(iii)) can unjam along the vertical, and this is an unstable mode of $\CS$. The fraction of 1-1 bonds are denoted by $\eta$.  We can quench further (protocol $\Qc$) by converting all the 1-1 to 0-0 bond pairs from configurations in $\CS$, and this would result in a completely jammed structure \cite{TorqJPCB2001} or an inherent structure~\cite{StillJCP1964}, see Fig 1(iv). We denote the space of completely jammed configurations by $\CJ$. The the jammed/inherent structure volume fractions is  $\frac{\pi d^2}{4h[\sqrt{(2d-h)h}(1-\theta+2\eta) + d(\theta-2\eta)]}$, where $\theta$ is the fraction of 1-bonds. The closed pack structure for this model has an arrangement of only 0-bonds. 
 
{\it Density landscape}~\cite{StillJCP1964} for $\q$, is a manifold in the $DN+1$ dimensional space constituting the $D N$ dimensional configurational space $\CR$ and the inverse of the fluid volume fraction $\phi^{-1}$. The density landscape of $\q$ can be divided into basins: configurations in $\CR$ that are mapped to the same configuration in $\CJ$ through a quench ($\Qa\Qb\Qc$). Equilibrium compression of the fluid results in the fluid exploring deeper regions (larger $\phi$)  of the landscape. In this letter, we identify the {\it onset volume fraction} $\phi_o$, beyond which there is a steep increase in the depth of the $\CJ$'s explored and the dynamics changes from free diffusion to {\it landscape influenced} as seen in generic glassformers~\cite{SastryNature1998}, in this regime the dynamics are characterized as non-Arrhenius (for fragile glassformers). The landscape can alternatively be divided into regions of saddle basins, where the basin is now defined as a collection of configurations in $\CR$ that are mapped to the same configuration in $\CS$ through the quench $\Qa\Qb$. Configurations in $\CS$ are characterized by the number of unstable modes $\eta$.  Computational studies\cite{AngelaniPRL2000, BroderixPRL2000, CoslovichArxiv2018} on the Kob-Anderson glass former\cite{KobAndersen1995} shows that as the temperature is lowered, the fraction of unstable modes decreases and almost vanishes resulting in a  dynamical cross-over from non-Arrhenius dynamics (fragile) to Arrhenius dynamics (strong).  In $\q$ model, the dynamical cross-over occurs at $\phi_d$, $\eta$ and $\phi_d$ had been calculated earlier~\cite{MahdiPRE2015}. Across $\phi_d$, simulations show a fragile ($\phi <\phi_d$) to strong ($\phi>\phi_d$) cross-over~\cite{MahdiPRL2012}. \\
{\it Geometry and Landscape}: First, we show the relationship of a configuration in $\CR$  with the corresponding configurations in $\CT$, $\CS$ and $\CJ$ in $\q$. We then connect the dynamics of $\CR$ to the dynamics in  $\CT$ to construct a KCM.  \\ 
Consider a low $\phi$ configuration in $\CR$. A typical configuration in $\CT$ would look like Fig.2(A)(i).  Here, fluid can access all $\CJ$ configurations when quenched. Let the mean horizontal distance between the neighboring discs in $\CR$ and $\CT$ are $u_C(\phi)$ and $u_T(\phi)=\braket{\Delta_x}$ respectively. If we were to compress a low $\phi$ configuration in $\CR$ quasi-statically and simultaneously monitor the corresponding configuration in $\CT$, we would come across a $\phi=\phi_o$, wherein the disks in $\CT$ experience a ``caging" constraint arising from the walls, as shown in Fig.2(A)(ii).  Under this condition, the center of the middle disc (orange) is coincident with the line segment connecting neighboring centers, while the neighbors touch the opposite walls. This occurs when $u_T(\phi_o)=u_T^*=\frac{1}{2}\sqrt{4d^2-(h-d)^2}$ at $\phi=\phi_o=0.390$. This constraint on the configurations in $\CT$, in turn, restricts the quench $\Qb\Qc$ from sampling low volume fraction with configurations with $u_T>u_T^*$ in $ \CJ$ resulting in a steady increase in $\theta$ of the explored configurations in $\CJ$ beyond the so-called onset volume fraction $\phi_o$. Fig.2A(i) and (iii) show the typical local geometry of configurations in $\CT$ below and above $\phi_o$  respectively, and 2(B) shows $u_C$ and $u_T$ {\it vs} $\phi$.  In Fig 2(C), we see a rapid decrease in $\theta-2\eta$ (implying a rapid increase in inherent structure volume fraction) beyond $\phi_o$. At $\phi=\phi_d=0.467$, $u_C(\phi_d)=u_T^*$, the disc movements in $\CR$ is restricted by entropic barriers, resulting in Arrhenius dynamics. Consequently, the number of unstable modes $\eta$ is very small beyond $\phi_d$ (Fig. 2(D)). One calculates $u_c=\frac{1}{N}\frac{\partial lnZ}{\partial P_L}$.
$u_T$,  $\theta$ and $\eta$ are calculated from the coarse-grained partition functions $G^{(3)}_1, G^{(4)}_1, G^{(5)}_1$  defined below:
\begin{equation}
	G^{(m)}_1(y_1,y_m)=\int e^{-\gamma \Delta_m}\prod_{i=1}^{m-1}K(y_i,y_{i+1})\prod_{j=2}^{m-1} dy_j
\end{equation}	
\begin{equation}
	u_T = \frac{1}{6}\mathcal{D} G^{(3)}_N;\theta=\frac{1}{4}\mathcal{D}G^{(4)}_{N};\eta=\frac{1}{5}\mathcal{D}G^{(5)}_N
\end{equation}	

Here the operator $\mathcal{D}=\lim_{\gamma\rightarrow 0}\frac{\partial}{N\partial\gamma}\ln Tr$.
The integration over $y's$ is from $d/2$ to $h-d/2$ and  $G^{(i)}_N(y,y'')=\int G^{(i)}_{N-1}(y,y')G^{(i)}_1(y',y'')dy'$ for $i=3,4,5$. Here, $\Delta_3=\Delta_x(y_1,y_2) +\Delta_x(y_2,y_3)$ and $\Delta_4$ , $\Delta_5$ are defined with other details in~\cite{MahdiPRE2015}. 
 $\phi_o$ and $\phi_d$ divide the landscape into three distinct dynamical regime which are characteristic of fragile glass-formers ~\cite{PabloNature2001}: (i) non-glassy regime (NGR):  when $\phi < \phi_o$ (called the diffusive regime for generic glass formers), (ii) landscape influenced regime (LIR):   $\phi_o<\phi<\phi_d$ and (iii) Arrhenius regime (AR) $\phi>\phi_d$. The identification of $\phi_o$ through $u_T$ links the average geometry of discs in $\CT$ to the landscape, see Fig. 2(B-D).
\begin{figure}[t]
\includegraphics[width=2.8in]{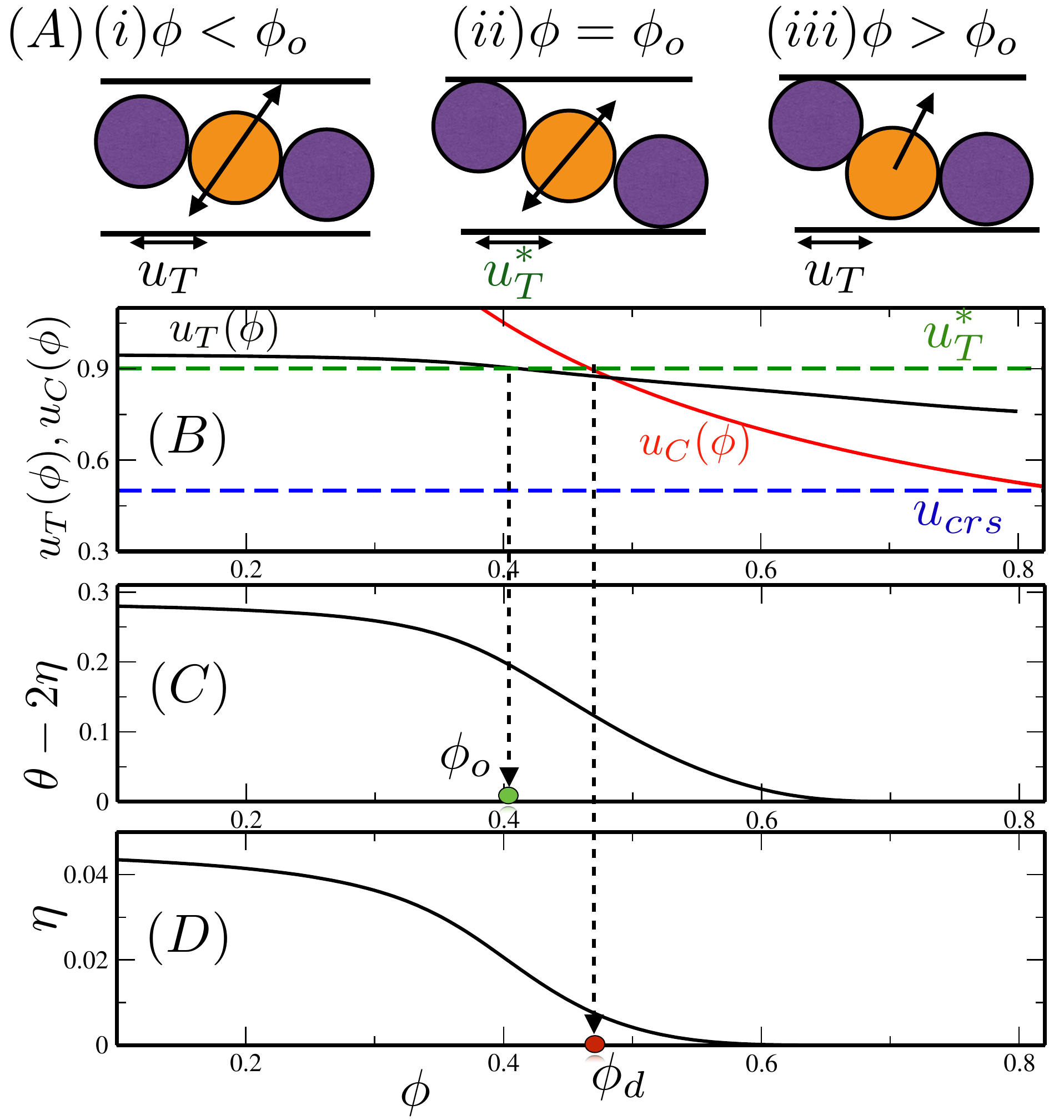}
\caption{  (A) Typical configuration in $\CT$ for volume fraction in $\CR$ at $\phi<\phi_o,\phi=\phi_o,\phi>\phi_o$. $\phi_o=0.390$, $\phi_d=0.467$, $\Delta_y^*=\Delta_y(\phi_o)$. (B)$u_T,u_C$ {\it vs} $\phi$. $u_T(\phi_o)=u_T*$ (C) $\theta-2\eta$ {\it vs} $\phi$. (D)$\eta$ {\it vs} $\phi$.}  
\label{fig2}
\end{figure}

 {\it KCM}: 
 We have a successive coarse-graining of configurations from $\CR\rightarrow\CT\rightarrow\CS$. With this in mind, we can construct the KCM. Consider a configuration with only 0-bonds (the closed pack configuration) in $\CS$, the centers of the discs in this configuration form a lattice, as shown in Fig. 3(a). Any configuration in $\CS$ can be represented on this lattice, by filling the appropriate lattice point with discs. One could think of dynamics on this lattice with discs moving only diagonally to the closest vacant position.  Fig.3(b) and (c) are typical configurations found in $\CS$. Shown Fig.3(d) is a special case when two vacant lattice points are neighbors resulting from the above dynamics; this is not a valid $\CS$ configuration since this would not be allowed by the $Q_{CT}Q_{TS}$ quench. The occurrence of this special case is rather rare \cite{rare}.  We introduce a space $\CSt$, which consists of all configurations in $\CS$ and all the special cases. A configuration in $\CSt$ can be mapped to a one-dimensional lattice $\Ld$ (Fig. 3(b-d). The discs are mapped to particles (white) and vacant sites to holes (black) on $\Ld$. The particles can move to the closest hole on $\Ld$.  We use transition rates from $\CT$ for the dynamics in $\Ld$ in contrast to the conventional transition state theory (TST), which uses $\CJ$~\cite{MoorePRE2014}. The above defines the KCM on $\Ld$. Transition rates from $\CT$ ensure that the KCM dynamics on $\Ld$ is very close to the dynamics in $\CR$. 

The average distance  between the adjacent discs along the y-axis in $\CT$ is  $\braket{\Delta_y}=\sqrt{d^2-u_T(\phi)^2}$.  $\braket{\Delta_y}$ is also the distance a disc needs to move to change the bond type. A new bond creation has an associated time scale of $\sim\braket{ \Delta_y}^2/k_BT$.
We use this as the time scale for the disc movement to the closest vacant lattice point when $\phi<\phi_o$.  For, $\phi>\phi_o$, the discs experience an entropic barrier $\sim e^{-\phi P_L\beta [u_{T}^{*}-u_T(\phi)]}$ in order to create a new bond. In $\CSt$, a particle can move when it is participating in a 1-0  bond (at a rate $p$) or a 1-1 bond (rate $q$). 
 On $\Ld$ (See Fig. 3) the rate of transition $p$ would correspond to a situation when a hole (black) moves to a particle (white) position, and the particle has no other hole as its neighbor on either side (Fig. 3(b),(c)).  A  hole transits to a particle position at rate $q$ when  (i) it has another hole as its neighbor, or (ii) the particle position it transits to has another hole as its neighbor (shown in Fig 3(c),(d)). The rates of transitions based on the three distinct dynamical regimes are then:
\begin{equation}\label{rates}
\text{rates} =
  \begin{cases}
    p(\phi)=\frac{k_BT}{\braket{\Delta_{y}}^2}   &  ~\phi\le \phi_o \\
    p(\phi)=e^{-\phi P_L\beta [u_{T}^{*}-u_T(\phi)]}\frac{k_BT}{\braket{\Delta_{y}}^{*2}}  & \text{if } \phi>\phi_o\\
    q(\phi)= \frac{k_BT}{\braket{\Delta_{y}}^{2}}  & \phi< \phi_o\\
    q(\phi)= \frac{k_BT}{\braket{\Delta_{y}}^{*2}}  & \phi> \phi_o\\
  \end{cases}
\end{equation}

{\it Relaxation and fragility}: Consider equilibrium configurations of $N$ particles on $\Ld$ with $L$ sites. At some instance, $t=0$ we define all the particles as not having participated in a transition. Let $R(t)$ be the number of particles which have not moved after time $t$.  
\begin{figure}[t]
\includegraphics[width=3.2in]{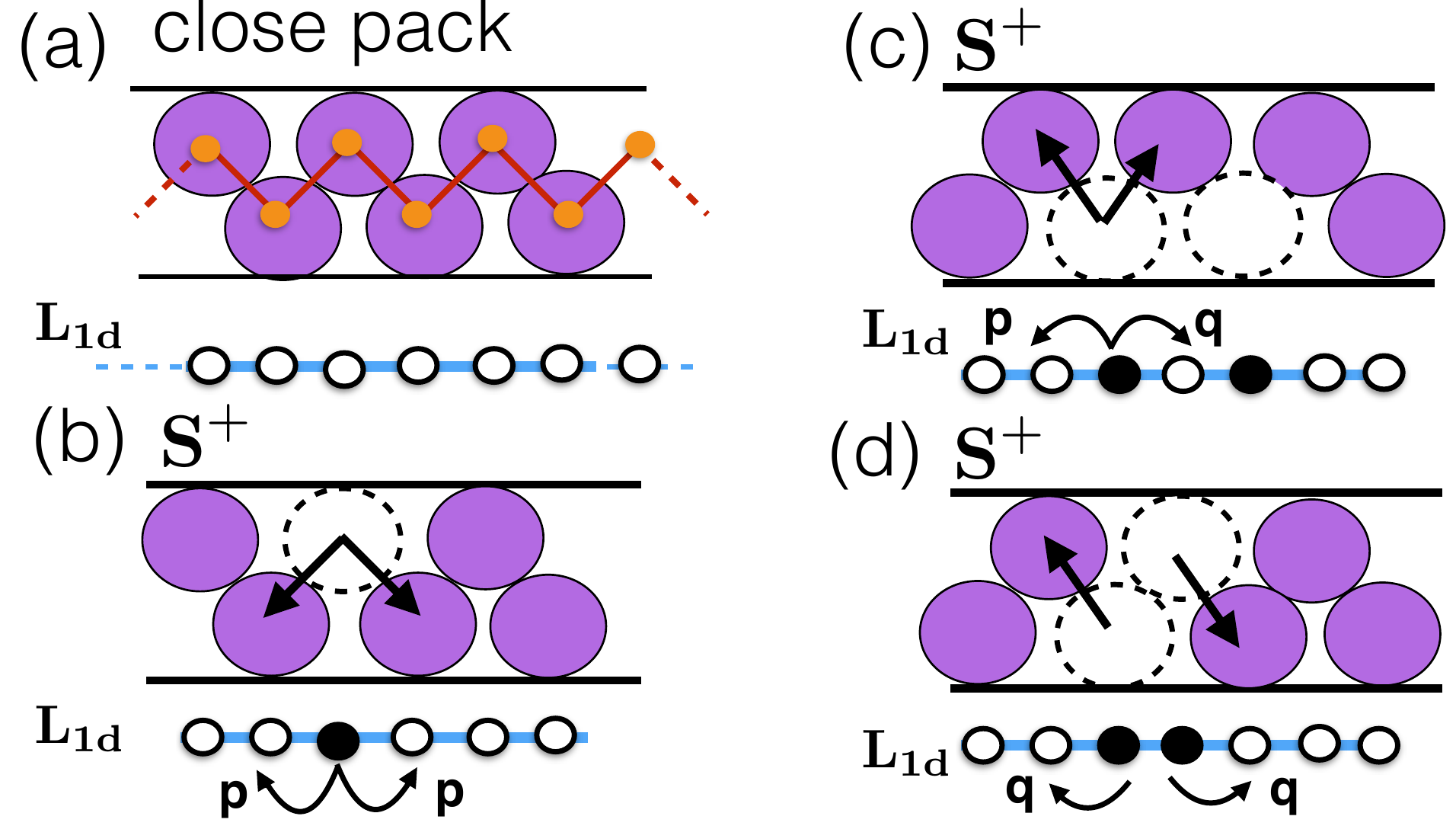}'
	\caption{(a) Close pack disc centers define a lattice (red edges-orange vertices), this lattice is mapped to a 1d lattice $\Ld$ shown below. (b) and (c) are typical configurations in $\CS$, the holes are black, and particles are white on $\Ld$. (d) Special configuration arising from dynamics. $\CSt$ is defined to include configurations such as (b),(c), and (d). $\CSt$ is mapped to $\Ld$.The rates p and q are defined in \eqref{rates}. }  
\label{fig3}
\end{figure}
 The fraction of particles participating in transitions with rate $p$ (1-0 bond pairs) for a given $R(t)$ is
 $(\theta-2\eta)R$ and those with rate $q$ (1-1 bond pairs) is $\eta R$. The master equation for this pure death process is:
\begin{equation}
	\partial_t P_t(R)=[p(\theta-2\eta) + \eta q][(R+1)P_t(R+1)-R P_t(R)]  
\end{equation}
with boundary conditions at $t=0$, $R=N$. One obtains, $<R(t)>=\sum_{R-0}^{R=N}RP(R,t)=N \exp(-t/\tau)$. 
The relaxation time is 
 \begin{eqnarray}\label{taueqn}
 \tau =  \frac{1}{[p(\theta-2\eta)
                                     +q\eta]}
 \end{eqnarray}
Arrhenius regime $\phi>\phi_d$ , the saddle point index $\eta\rightarrow0$, eq. \eqref{taueqn} $\tau\rightarrow \exp[P_L(u_T^{*}- u_{T})]/\theta$ thus the dynamics is strong since $\theta$ has a weak dependence of $\phi$ compared to the exponential numerator. In the ``diffusive" limit $\phi<\phi_o$, $p,q\rightarrow k_BT/\braket{\Delta_y}^2$ and  eq.\eqref{taueqn} results in $\tau=\frac{\braket{\Delta_y}^2}{\theta k_BT}$. Using transition state theory (TST) relaxation in the dense regime was calculated by Godfrey {\it et. al} ~\cite{MoorePRE2014} for $\q$ (see Fig. 4). In Fig. 4 (inset), we calculate the fragility $f=\frac{d log_{10}\tau}{d\phi PV/Nk_BT}$. We find a dynamical transition at $\phi_o$, with first order features (shown in purple ellipse). Interestingly, the s-ensemble approach also predicts a first order dynamical transition in KCMs \cite{Garrahan2007,Garrahan2009}. Though it is tempting to identify them as the same, this  needs further investigation. 
\begin{figure}[t]
\includegraphics[width=3.5in]{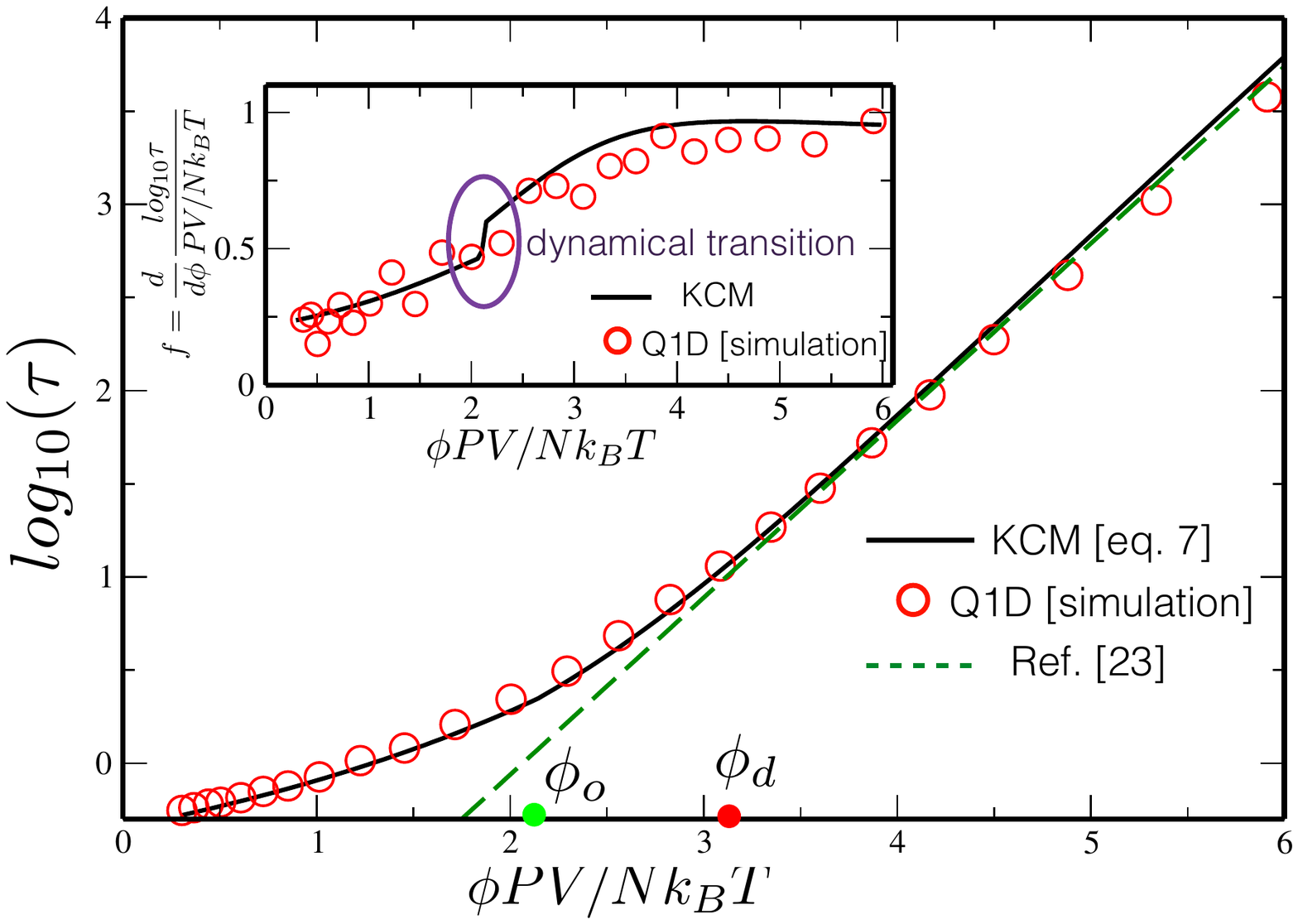}
\caption{Relaxation time derived for the KCM (black). Relaxation times of the $\q$ using molecular dynamics \cite{MahdiPRE2015} (red circles). Godfrey {\it et al.}\cite{MoorePRE2014} have calculated the relaxation times (green dashed) in the dense regime using inherent structure transition.[Inset] Fragility index of the KCM across the fragile-strong cross-over (black), fragility index of the $\q$ (red squares) \cite{MahdiPRE2015}. Dynamical transition with first-order features at $\phi_o$ is marked using an ellipse }  
\label{fig4}
\end{figure}

{\it Glassy influence on TNPF}:
Isolating the glassy contribution to correlation function in fluids has been a challenge. KCM on $\Ld$ is a variant of the simple symmetric exclusion process (SSEP). TNPF for the SSEP can be calculated exactly via the Bethe ansatz (BA)~\cite{Domb2000} and correlation functions obtained from it upon integration. The approach to equilibrium in the three dynamical regimes using the TNPF is our interest. The BA approach helps us isolate the influence of glassy dynamics on the  TNPF  $P(X^t|X^{0})$ (where, $X^t=\{x_1...x_N\}$ are the position of the holes on $\Ld$ at time $t$, given they were at $X^0= \{x_1^{0},..x_N^{0}\}$ at $t=0$).  We begin by writing 

\begin{equation}\label{P0}
P(X^t|X^{0})=\int_{0}^{2\pi}..\int_{0}^{2\pi}e^{-E t}\psi(X^t)\prod_{i}e^{-ik_jx_j^{0}}dk_j
\end{equation}
 Here $\psi(\{x_j\})$ is the spatial part of the function~\cite{Domb2000}. The energy $E=\sum_i^N\epsilon_i$ and $\epsilon_i = p(1-cos(k_i))$. Though the solution is set in the complex space $\mathbb{C}^N$, it is interpreted as a probability only on the real axis. The 2-hole probability function when $x_1<<x_2$ satisfies the master equation:
\begin{eqnarray}\label{c1} 
&\partial_t \psi(x_1,x_2)& = p \psi(x_1-1,x_2)+ p \psi(x_1,x_2-1) \\\nonumber
                  &+& p \psi(x_1+1,x_2) +  p \psi(x_1,x_2+1))-4 p \psi(x_1,x_2)\nonumber
\end{eqnarray}
 A boundary like condition is given by the non-allowed transitions :
 \begin{eqnarray}\label{c2}\nonumber
 &p& \psi(x+1,x+1) + p\psi(x,x)+(p-q) (\psi(x-1,x+1)\\
 &+& (p-q) \psi(x,x+2)-2 (2p-q) \psi(x,x+1)=0
 \end{eqnarray}
Using the  Bethe ansatz  :$ \psi(x_1,x_2)=\Aotz +\Atoz$ and \eqref{c1}and \eqref{c2}, with $z_i=e^{ik_i}$, $w=p-q$ we find the scattering function $S_{21}$ to be:
 \begin{equation}\label{Seqn}
S_{21}=\underbrace{1-\overbrace{2\frac{(z_1+z_2)}{1+z_1z_2-2z_1}}^{H^I}}_{SSEP} 
+\underbrace{\overbrace{w\frac{(z_2-z_1)}{z_1(w+p z_2)} \frac{(1+z_1z_2-2z_2)}{(1+z_1z_2-2z_1)}}^{H^G}}_{Glassy}
\end{equation}
In $S_{21}$, unity is the non-interacting part, and  $H^I$ is the interaction term of SSEP. The term  $H^G$, only arises beyond the glassy onset $\phi>\phi_o$, since $w= 0$ for $\phi\leq\phi_o$, and this is the glassy contribution. 
In the (i) NGR:  the scattering matrix and the BA conditions with $w=0$ results in a TNPF which is precisely that of SSEP \cite{Domb2000}. (ii) LIR: For 3-hole collisions, the scattering matrix factorization fails. Fortunately, in the LIR, $\eta<<1$ even at $\phi\sim\phi_o$ and gets steadily smaller with increasing $\phi$, making two-hole collisions sparse and hence three-hole collisions significantly rare. We could thus approximately write  $\psi(\{x_j\})$ in terms of just the two-body scattering terms :
\begin{equation}
\psi(\{x_j\})\approx z_1^{x_1}z_2^{x_2}..z_N^{x_N} + \sum_{i\neq j} S_{ij} z_1^{x_1}..{z_i}^{x_j} ..{z_j}^{x_i}..z_N^{x_N}
\end{equation}
Using equations \eqref{P0} and \eqref{Seqn} and with $2\rightarrow i,1\rightarrow j$ in Eq.\eqref{Seqn}, using the modified Bessel function: $I_n(t)=\frac{1}{2\pi}\int_{-\pi}^{\pi} d\phi \exp(i\phi n + t cos(\phi))$, $t\rightarrow \infty$ with $n^2/t$ held constant, we have the identity~\cite{Shutz2001,Magnus1987}:
\begin{equation}
e^{-t}I_n(t)\sim\frac{1}{\sqrt{2\pi t}}\exp(-n^2/(2t))
\end{equation}
and we find:
\begin{eqnarray}\label{P1}
&&P(X^t|X^{0})\sim  \underbrace{\prod_j^N e^{-pt}I_{x_j-x_j^0}(pt)}_{SSEP} \\\nonumber
&&+\underbrace{\prod e^{-pt}[I_{x_i-x_j^0}(pt)-2 \sum_{ij} g_{ij}(pt)] }_{SSEP(1-H^I)}\\\nonumber
&&+\underbrace{w\sum_{ij}h_{ij}^{(1)}(pt)+h_{ij}^{(2)}(pt)-2h_{ij}^{(3)}(pt) }_{Glassy(H^G)}\\\nonumber
\end{eqnarray}
Here, $g_{ij}(pt)=\sum_{mn} I_{(m+1)x_j-x_j^0}(pt)I_{(m+n)x_i-x_i^0}(pt)$ and $h_{ij}^{(1)}(pt),h_{ij}^{(2)}(pt),h_{ij}^{(3)}(pt)$ are defined in \cite{refeq}.
In the (iii) AR: $\eta\approx 0$, this implies that the hole-hole interaction is negligible, hence 
$\psi(x_1..x_N)=1$ using \eqref{P0},  thus the TNPF reduces to:
\begin{equation}\label{P2}
P(X^t|X^{0})\sim \prod_{i}^N I_{x_i-x_i^0}(pt)
\end{equation}
This establishes TNPF in the three distinct dynamical regimes which are Generic to glass-formers. Equation \eqref{taueqn} provides us with an analytically transparent relationship between thermodynamics and the KCM relaxation time. The relaxation time in this model does not follow Adam Gibbs~\cite{MahdiPRE2015} or the parabolic law developed by Elmatad, Chandler and Garrahan~\cite{ECG2009,ECG2010}. It would be interesting to see if \eqref{taueqn} is consistent with the parabolic law in KCMs studied in ~\cite{ECG2009,ECG2010}. We note that the relaxation time in this model is not controlled by a growing static length scale but by local entropic barriers trivially consistent with Wyart and Cates ~\cite{WyartCates2017} scenario. Further, it is intriguing to note that both the partition function and the s-ensemble ~\cite{Garrahan2007, Garrahan2009} approach indicate a first-order dynamical transition, whether the transition in the s-ensemble approach is a signature of the glassy onset is a question which needs further investigation.

I would like to thank  Vinod Krishna, Bhaswati Bhattacharyya, Masaki Sasai,  and Srikanth Sastry for their comments.
This work is supported by Japanese Science and Technology Agency  grant JST CREST 446 JPMJCR15G2.


\begin{thebibliography}{999}
\bibitem{rare} A transition from 1-1 bond is required to create this situation and as shown in Fig 2(D), The fraction of 1-1 bonds $\eta<0.04$ even at the lowest $\phi$.
\end{thebibliography}
\end{document}